*Article*

# Enhancing dysarthria speech feature representation with empirical mode decomposition and Walsh-Hadamard transform

**Ting Zhu** [1], **Shufei Duan** [1,*], **Camille Dingam** [1], **Huizhi Liang** [2], **Wei Zhang**[3]

[1] College of Electronic Information and Optical Engineering, Taiyuan University of Technology, Taiyuan, Shanxi 030024, China; zt_april@163.com (T.Z.); dingamcamille8@gmail.com (C.D.)
[2] School of Computing, Newcastle University, Newcastle NE17RU, UK; huizhi.liang@newcastle.ac.uk
[3] Taiyuan Hospital, Peking University First Hospital, Taiyuan, Shanxi 030032, China; zwwpa@163.com
* Correspondence: duanshufei@tyut.edu.cn (S.D.)

**Abstract:** Dysarthria speech contains the pathological characteristics of vocal tract and vocal fold, but so far, they have not yet been included in traditional acoustic feature sets. Moreover, the nonlinearity and non-stationarity of speech have been ignored. In this paper, we propose a feature enhancement algorithm for dysarthria speech called WHFEMD. It combines empirical mode decomposition (EMD) and fast Walsh-Hadamard transform (FWHT) to enhance features. With the proposed algorithm, the fast Fourier transform of the dysarthria speech is first performed and then followed by EMD to get intrinsic mode functions (IMFs). After that, FWHT is used to output new coefficients and to extract statistical features based on IMFs, power spectral density, and enhanced gammatone frequency cepstral coefficients. To evaluate the proposed approach, we conducted experiments on two public pathological speech databases including UA Speech and TORGO. The results show that our algorithm performed better than traditional features in classification. We achieved improvements of 13.8% (UA Speech) and 3.84% (TORGO), respectively. Furthermore, the incorporation of an imbalanced classification algorithm to address data imbalance has resulted in a 12.18% increase in recognition accuracy. This algorithm effectively addresses the challenges of the imbalanced dataset and non-linearity in dysarthric speech and simultaneously provides a robust representation of the local pathological features of the vocal folds and tracts.

## 1. Introduction

Dysarthria, also known as neurological speech disorder [1], occurs when muscles controlling vocal cords, vocal tract, mouth or other acoustic organ functions become paralyzed or incoordination [2]. This impacts rhythmic accuracy during speech. It is commonly seen in conditions such as brain injury and other diseases [3]. Dysarthria is categorized as MA80 speech disorder in the International Classification of Diseases 11th Revision (ICD-11). This reflects abnormalities in the amplitude, speed, and stability of respiratory and phonological movements during speech production in patients with dysarthria. Speech therapy is one of the common therapeutic means [4]. Studying pathological speech helps speech therapists to carry out systematic assessment and rehabilitation. Among them, the extraction of acoustic features with good pathological differentiation is the key to voice pathology detection (VPD) systems. The common acoustic features of pathological speech are pitch, Mel Frequency Cepstrum Coefficients (MFCC), Linear Prediction Coefficients (LPC) and so on. Some researchers treated the speech production system as a linear model and extracted various features to research the regularity of pathological voice and its difference from normal, or to detect the severity of dysarthria [5-7]. The aforementioned studies have demonstrated the effectiveness of traditional acoustic features for VPD. Speech is viewed by the linear model as the result of a source filter system. Pre-processing operations like framing, windowing are often performed on feature extraction to divide speech signals into short-time stationary segments due to the non-stationary nature of speech signals. Although this kind of processing can capture the time-varying traits of speech to a certain extent, it may cause problems such as signal loss and discontinuity between frames. Therefore, researchers need to consider the nonlinear and non-stationary characteristics of the signals more thoroughly to better understand and utilize their information.

Given the mechanism of signal generation as well as the nonlinear nature of the speech system, speech signals are nonlinear [8]. It is evidenced by the linear distortion during speech transmission, nonlinear filtering of the vocal site and nonlinear perception of the human ear auditory system in the signal extraction process. At the same time, speech exhibits non-stationary characteristics [9] attributable to dynamic variations in articulatory configurations, such as lip shaping and respiratory modulation. Dysarthria patients, with organic impairment of vocal folds or mouth cavity, may demonstrate pronounced nonlinear vocal vibrations. Neglecting such chaotic phenomena risks compromising model performance and reducing the ability to effectively distinguish pathological from normal speech. Consequently, researchers aim to design nonlinear models framing speech generation as a nonlinear dynamical system influenced by chaos theory. Vieira et al [10] proposed non-stationary indices as an adaptive segmentation method for pathological speech, which improved the accuracy of VPD by 18% compared to linear systems. Gour et al [11] extracted the atypical parameters to use support vector machines for disease classification, with 73% recognition accuracy. These findings underscore the imperative of selecting appropriate nonstationary and nonlinear processing techniques for the precise extraction of feature information from pathological speech signals.

In the studies of dysarthria speech, it has been found that articulatory impairments are reflected in the vocal tract and vocal folds containing a considerable amount of pathological information. However, the commonly used acoustic features tend to only reflect single-sided information. For example, MFCC, LPC, and formant mainly manifest the information of the vocal tract, while fundamental frequency indicates vocal fold motion. Based on this, seeking for feature parameters that can more comprehensively profile the VPD system is the focus of this study. Empirical Mode Decomposition (EMD), an adaptive time-frequency signal analysis method [12], can resolve the local information of the original signal in different time scales, and accurately express the instantaneous frequency dynamics [13, 14]. EMD boasts strong time-frequency focus and signal-to-noise ratios, which render it particularly suitable for non-linear, non-stationary signals. For example, Krishnan [15] employed EMD to extract nonlinear speech signal features for the recognition of seven different emotions. Rueda [16] applied the binary filter bank characteristic of EMD to identify Parkinson speech. Zhang et al [17] combined EMD and spectral analysis to distinguish between Parkinson patients and healthy people. Hammami [18] introduced an advanced EMD-Discrete wave transform analysis for higher-order statistical feature extraction in pathological speech. Our work will leverage the ability of EMD to adaptively decompose dysarthria speech. Capitalizing on the proficiency of EMD to adaptively decompose speech signals, our work aims to construct a more comprehensive framework for capturing pathology information. It is specifically tailored to encapsulate the local characteristics of the vocal tract and vocal folds, thereby transcending the limitations of existing methodologies in profiling dysarthric speech.

Fast Walsh-Hadamard Transform (FWHT) is a fast algorithm for solving linear differential and integral equations [19]. By decomposing signals under square waves with different oscillation frequencies, it can be used to quickly analyze signals, image, and is ideal for signal transformation [20]. Subathra [21] used it to decompose EEG signals and extracted a series of nonlinear features; Hela et al [22] applied WHT with singular value decomposition to medical watermarking images and obtained robust results against noise. Mohsen et al [23] analyzed EEG of epileptic with the help of FWHT technique, which helps to effectively deduce seizure EEG from non-epileptic. However, existing research on FWHT has mainly focused on image processing, and FWHT has not yet fully been explored to analyze and study pathological speech signals. The time complexity of the algorithm is much lower than that of the FFT, which is highly efficient for processing data in medical fields such as pathological speech. Furthermore, pathological speech signals are usually marked by high spectral energy concentrated in a few frequency components. FWHT, as a frequency-domain processing method, is helpful for analyzing the abnormal frequency components in voice pathology. Meanwhile, its information compression and parallel computing capability can effectively improve the transmission efficiency and processing speed. Hence, this paper innovatively introduces this algorithm to deal with the abnormal fluctuations of pathological speech signals, which further accelerates the reconstruction speed of the signals, and can adaptively construct the speech enhancement characteristics of the dysarthria.

In light of the comprehensive analysis above, there are two shortcomings for VPD systems: 1) the problem of non-stationary and non-linear characteristics of pathological speech and 2) the limitations of the current lack of feature parameters that simultaneously portray the local pathological information in the vocal folds and vocal tracts of dysarthria patient. This paper introduces the application of FWHT and EMD to enhance the detection and capture of the pronounced non-linear and attributes associated with the aberrant muscular movements resulting from dysarthria. An adaptively dysarthria speech feature representation algorithm (WHFEMD) is proposed to further enhance the dysarthria speech recognition system.

**2. WHFEMD methodology**

The overall diagram of the pathological speech recognition method proposed in this paper is shown in Figure 1, which mainly includes the original pathological speech signal, speech processing and decomposition based on EMD and FWHT, feature extraction, and classification. The proposed algorithm in this paper is named WHFEMD, where FEMD (Fast Fourier Transform Along with Empirical Mode Decomposition) refers to the step of processing speech into FFT and EMMD. Specifically, EMD is used to extract the non-stationary and non-linear characteristic information in the original speech as well as the local feature information in the vocal tract. FWHT is then applied to derive the transformation coefficients. This step aims to improve the anti-noise performance and the calculation speed and to conduct speech enhancement of dysarthria to achieve the purpose of pathological speech feature enhancement. Finally, classification algorithms recognize and grade the level of dysarthria based on the enhanced features. This complete process labeled WHFEMD is designed to strengthen feature learning from raw dysarthria speech recordings.

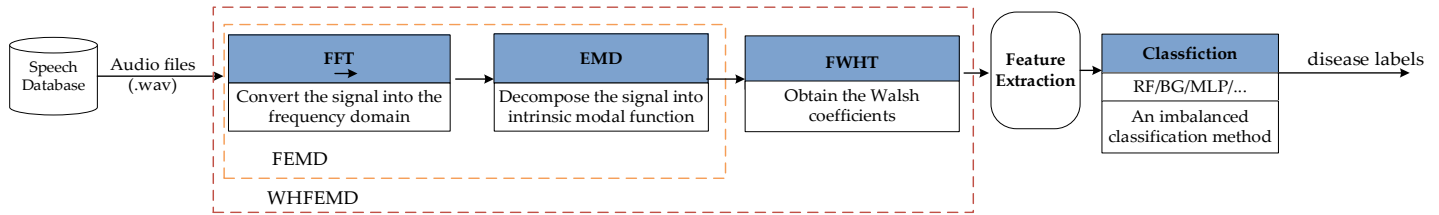

**Figure 1.** The diagram of the proposed pathology speech recognition system.

*2.1. Empirical Modal Decomposition*

EMD is an adaptive, data-driven decomposition technique that stepwise breaks down signals based on their inherent time-scale dynamics. This process obtains several intrinsic mode functions (IMFs) and a residual component representing the signal. For a component to qualify as an IMF, it must satisfy two criteria:
- The number of extreme and zero-crossing points can differ at most by one.
- The mean of the upper and lower envelopes defined by the local maxima and minima respectively is zero at any given time.

Assume that given a signal $x(t)$, its EMD is defined as:

$$x(t) = \sum_{i=1}^{n} IMF_i(t) + r_n(t) \tag{1}$$

Where $IMF_i(t)$ is the intrinsic mode function of the signal and $IMF_i(t)$ is the residual.

The main advantage of introducing EMD decomposition of dysarthria speech in this paper is that the IMFs serve as the basic components of signals. This localized, data-driven approach is well-suited to irregular pathological waveforms. The IMFs contain all the formant frequencies, vocal folds and vocal tract information that are closely related to speech. From the signal processing point of view, high-frequency components mainly reflect the instantaneous changes and fluctuations of the signal, then the standard deviation in the high-frequency IMFs may be more sensitive. Meanwhile, variances in lower frequency IMFs could reveal longer-term trends and shifts. In short, the EMD decomposition affords a highly nuanced view of both normal and disordered speech production mechanisms, capturing pathological traits that distinguish patient groups. This detailed portrait enhances our ability to analyze and understand dysarthria.

*2.2. Fast Walsh-Hadamard transformations*

For a given signal $x(t)$ and its eigenmode function $IMF_i(t)$, its FWHT transform coefficients can be defined as:

$$FWT(u) = \frac{1}{N} \sum_{u=0}^{N-1} IMF_i(t) \prod_{k=0}^{j-1} (-1)^{b_k(u)b_{(j-1-k)}(u)} \tag{2}$$

Where $FWT(u)$ denotes the Walsh-Hadamard transform coefficient, $b_k(u)$ represents the value of the $k+1$th bit of the binary value of $u$, $b_{(j-1-k)}$ is the value of the $j-k$ bit, $u=0,1,2,…,N-1$ denotes the current $u$-th data point, $N$ means the number of sampling points, generally an integer power of 2, $j$ is the number of intrinsic modal functions, and $k$ denotes an integer.

The FWHT uses ±1 discrete values instead of the complex exponentials as in the FFT. And it also performs only additions and subtractions, conferring lower computational complexity and faster operation speed. In addition, FWHT compresses the energy such that the first coefficient usually contains most of the signal energy. This allows the first coefficient to largely represent the original signal.

Figure 2(a) displays the spectrum of the IMF for the first 5 orders of the dysarthria speech. It can be observed that the spectra of the IMFs reveal different speech components, with $IMF_1$ carrying almost all the energy of the signal. Its spectrum decreases approximately twice times with each successive IMF order. Figure 2(b) then shows its WHT coefficients. As the theory proposed, most energy is compressed into the first coefficient sequence. This validates that the FWHT inherently concentrates signal power in the leading terms, with $FWHT_1$ again dominating.

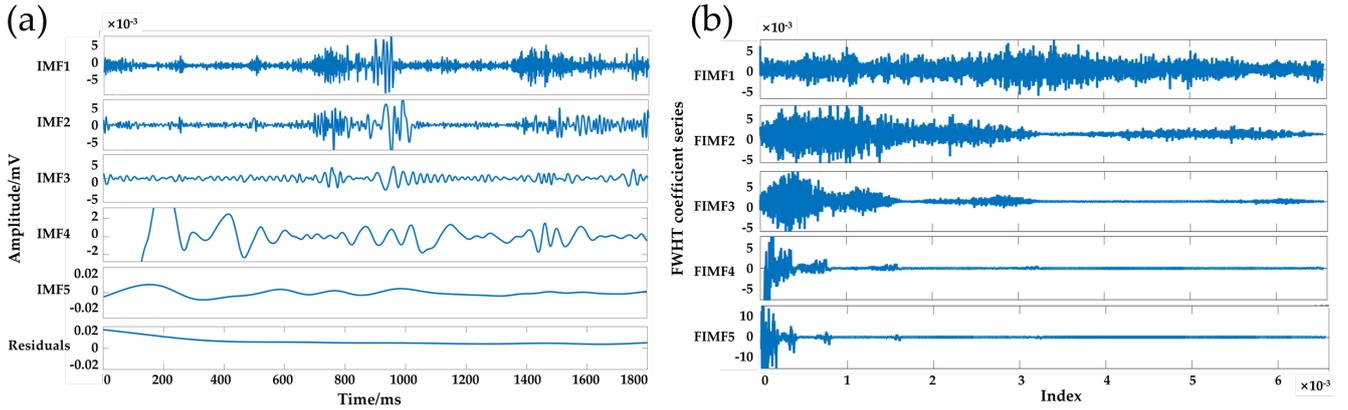

**Figure 2.** EMD results and WHT coefficients of dysarthria speech: **(a)** EMD decomposition of dysarthria speech (IMF1-IMF5); **(b)** FWHT coefficients for dysarthria speech.

*2.3. WHFEMD algorithm*

The WHFEMD algorithm is mainly divided into signal decomposition and feature extraction. Firstly, FEMD is obtained by processing the speech using FFT and EMD; then WHFEMD is obtained by FWHT transform, and the statistical features and common acoustic features of FEMD and WHFEMD are extracted as the data input for subsequent classification. The steps are shown below:

(1) Suppose that the magnitude spectrum of the original signal is obtained using FFT. The Fourier transform of the time signal and the magnitude spectrum obtained after FFT are $s(f)$ and $p(f)$, respectively.
(2) Introduce EMD on the FFT amplitude spectrum $p(f)$ to get the IMFs. It should be noted that EMD decomposition stops when the residual $r_j(f)$ of the $IMF_j(f)$ has become a monotonic function or constant. The specific function expressions are as follows.

$$P(f) = \sum_{i=1}^{j} IMF_i(f) + r_j(f) \tag{3}$$

Where $IMF_i(f)$ is the *i-th* intrinsic mode function decomposition, $i=1,2,3,…,j$, $r_j(f)$ denotes the average trend of the signal, which is a constant or monotonous sequence, and $j$ means the number of IMFs. In this paper, after a series of the severity of disease experiments, Table 1 shows the peak performance, with 71% accuracy on TORGO and 44% on UA Speech. So, the experimental part of this paper is selected as *j=5*.

**Table 1.** Selection of IMF series (integer recognition rate).

| IMF Series (j) | TORGO (%) | UA Speech (%) |
|---|---|---|
| 1~4 | 65 | 38 |
| 1~5 | 71 | 44 |
| 1~6 | 69 | 44 |
| 1~7 | 69 | 43 |
| 1~8 | 66 | 41 |
| 1~9 | 63 | 40 |
| 1~10 | 62 | 40 |

(3) The FEMD signal is constructed by applying FWHT to the first 5 orders of IMFs to obtain its transform coefficients.
(4) Extract the statistical features as well as the enhanced features of Power Spectrum Density (PSD) and Gammatone Frequency Cepstral Coefficients (GFCC) from the FEMD and WHFEMD.

*2.4. Feature extraction*
(1) Statistical features (FESF/WHFESF)

Since time domain waveforms differ greatly between dysarthria and normal speech, statistical features of IMFs were extracted from the FEMD and WHFEMD coefficients respectively, named as FESF and WHFESF, to form $j \times a$ dimensional feature vectors, where $j$ represents IMF order and $a$ is the number of statistics. The experiments in this paper take $a=5$, specifically the mean, standard deviation (SD), maximum, minimum and variance to obtain a 25-dimensional statistical feature. Variance reflects signal strength variation and temporal instability. SD focuses on deviation from the mean value, capturing fluctuation patterns. Previous work [24, 25] found dysarthria speech has a longer duration, reduced amplitude, slower speed and more irregular movement trajectories than normal. Therefore, maximum and minimum were extracted to characterize abnormal articulatory movement. Therefore, extracting the maximum and minimum features helps to capture the sound signal characteristics associated with abnormal articulatory movements. And Z-score normalization method is used to ensure that sound signals are analyzed on a similar scale across different corpora, to reduce the impact of sound amplitude differences on the accuracy. It converts values into deviation scores relative to the sample mean and SD, so that the different corpus sound amplitudes exist as data values with similar distributions, thus reducing the variability. This method is still effective for capturing non-linear vibrations and anomalies in the speech of patients with different dysarthria. The RF classifier results show in Table 2 that the recognition rates on both databases improved by at least 0.49% and 0.52%. No performance reduction occurred due to strong variable correlations between SD and variance. Their use here for dysarthria speech captures the different characteristics and provides useful information for differentiating patients with dysarthria.

**Table 2.** Validation of variance and standard deviation.

| Dataset | Feature | Acc (%) | Feature | Acc (%) |
|---|---|---|---|---|
| TORGO | Only Std | 78.97 | Only Std | 80.58 |
| | Only Var | 78.90 | Only Var | 80.06 |
| | FESF | 79.45 | WHFESF | 81.10 |
| UA Speech | Only Std | 57.88 | Only Std | 62.45 |
| | Only Var | 57.39 | Only Var | 63.05 |
| | FESF | 58.37 | WHFESF | 63.99 |

(2) PSD Enhanced Feature (FEPSD/WH2FEPSD)

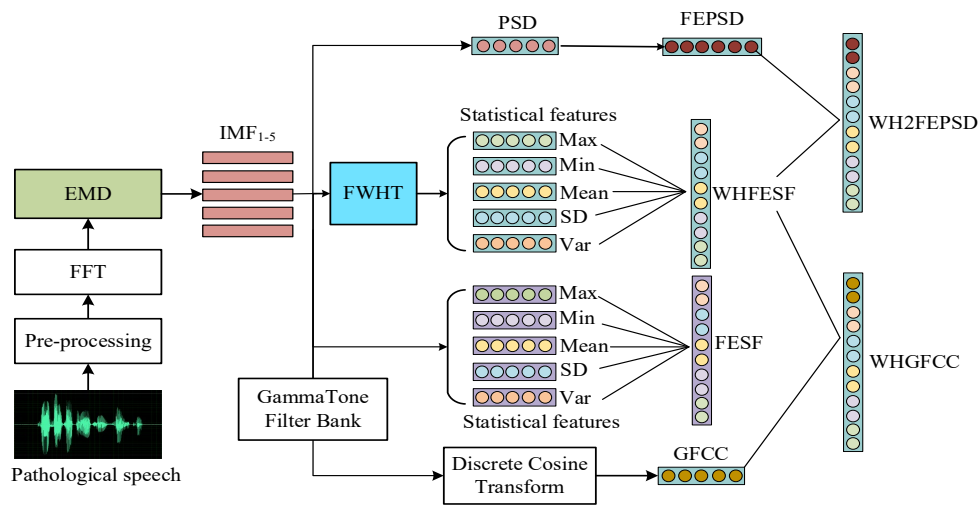

**Figure 3.** Flowchart of WHFEMD feature enhancement.

For random signals like speech, it is very effective to study its PSD. This paper uses the Welch method of the periodogram spectral estimation to extract the PSD of dysarthria speech. First of all, the input signal is segmented, then the FFT is applied to the segmented signal to estimate the PSD. These are averaged to obtain the final PSD. Segmentation length decides the spectral resolution and the overlap rate influences the degree of smoothing. The Welch method permits flexibility in adjusting the window size and overlap ratio. This enables fine estimation of local frequency information, handling non-stationary signals. In this experiment, each IMF is segmented and windowed via a 256-point Hanning window with 50% overlap. FEMD promotes the performance of PSD to obtain the new feature FEPSD. Additionally, WHFESF is combined with FEPSD to form the new feature WH2FEPSD (see Figure 3 for the process). After the FFT transform and EMD decomposition of pathological speech to obtain IMFs, the feature is mainly divided into: 1) the transform coefficients obtained from the FWHT transform of IMFs are evaluated to form the statistical feature set "WHFESF"; and 2) PSD of IMFs are obtained to form the feature set as FEPSD. Finally, these two feature sets are spliced and fused to obtain the new 30-dimensional feature WH2FEPSD ($j=5, a=5$). The researchers can change it according to the experimental need, so as to realize the purpose of feature dimension expansion. The function *welch()* is defined as:

$$P_{welch} = \frac{1}{T}\sum_{t=1}^{T} P_t(f) \qquad (4)$$

Where *T* is the number of segments of the segmented signal, and $P_t(f)$ is the periodogram of the *t-th* segmented signal.

(3) Enhanced features for GFCC (WHGFCC)

GFCC is a noise-robust enhancement of the widely used MFCC features for speech processing [26]. It simulates sharp basilar membrane filtering using GammaTone filters in equivalent rectangular bandwidths. While MFCCs provide high accuracy and low complexity, GFCC demonstrates greater noise immunity. Studies show it characterizes resistance to acoustic changes better than MFCC[27]. The extraction method is similar to that of MFCC, with the two main differences. One of them is the frequency scale, GFCC is based on the ERB scale, while MFCC is based on the Mel scale. It allows GFCC to provide higher resolution than MFCC in the low-frequency band [28]. On the other hand, GFCC applies exponential compression simulating auditory nonlinearity, and conferring better interference rejection than MFCC's logarithmic compression. To leverage these advantages, this paper combines GFCC with the statistical WHFESF features into a new parameter WHGFCC to optimize overall system test performance.

(4) Commonly used acoustic features for control experiments

To rigorously assess the performance of the proposed methodology, we have chosen some typical acoustic features for comparative evaluation, specifically including MFCC, LPC, GFCC, PSD, Pitch, and Line Spectral Pairs (LSP).

## 3. Experiments and results

### 3.1. Databases

UA Speech dataset [29]: created by University of Illinois with 15 cerebral palsy patients and 13 normal subjects. The corpus includes numbers, letters, computer commands, words, etc. In this paper, a total of 21,188 words from patients and normal people were selected.

TORGO dataset [30]: created by University of Toronto, it contains acoustic data, articulatory data from 8 patients with motor dysarthria (3 females, 5 males) and 7 normal (3 females, 4 males). The acoustic data recorded for short words and restricted sentences were selected for this study.

Both datasets allowed the study of dysarthria condition grading as shown in Table 3.

**Table 3.** Data selection on UA Speech and TORGO databases.

| The severity of dysarthria (UA Speech) | Samples | Total | The severity of dysarthria (TORGO) | Samples | Total |
|---|---|---|---|---|---|
| High | 2861 |  | Severe | 194 |  |
| Medium | 2280 |  | Moderate | 259 |  |
| Low | 2280 | 21188 | Mild | 596 | 4867 |
| Very Low | 3825 |  | Normal | 3818 |  |
| Normal | 9942 |  |  |  |  |

### 3.2. Experimental setup

In this paper, three classifiers from the integrated algorithm are used to accomplish the downstream classification task, namely Random Forest (RF), Bagging (BG) and Multilayer Perceptron (MLP). RF [29] uses both bagging and feature selection methods to improve accuracy without overfitting. The BG algorithm [31] generates a new resampled training set by randomly selecting replacement data from the existing training set. BG used in this work employs a decision tree post pruning algorithm. MLP [32] is a supervised neural network with input, hidden and output layers fully connected for classification problems. The datasets were divided into 70% training and 30% test sets. Feature representations from both traditional acoustic features (MFCC, LPC, PSD, Pitch, LSP) and the proposed algorithm were compared to validate effectiveness and reliability. All the experiments in this paper were conducted using an HP ZBook Power G7 mobile workstation running MATLAB 2022a. The processor configuration is Intel Core i7-11850H with 32GB of RAM.

### 3.3. Experimental results

This study employed a multi-pronged experimental approach to thoroughly evaluate the proposed feature enhancement methodology. First, WHFEMD and FEMD extracted statistical representations were compared against traditionally prominent acoustic features to benchmark initial classification performance. Considering inter-class imbalance is a critical issue, an adaptive unbalanced learning algorithm [31] was then introduced. This method comprehensively addresses imbalance by integrating PCA, SMOTE, and EM techniques specifically for discerning pathological speech severity. Applying this algorithm to the WHGFCC allowed further measuring algorithm flexibility and strength. Results on the TORGO and UA Speech databases achieved markedly classification beyond existing imbalance handling techniques.

3.3.1. Comparisons of commonly used acoustic features

On the UA Speech database (Table 4), WHFESF extracted via WHFEMD outperformed other features, with gains of 13.8, 12.65%, 7.47%, and 10.59% (RF), 12.16%, 8.62%, 7.38%, and 8.64% (BG), 11.72%, 9.05%, 7.6%, and 8.86% (MLP) compared to PSD, Pitch, LPC, LSP. The TORGO database saw similar results (Table 5), with WHFESF achieving 81.10%, 80.00%, and 78.84% accuracy for RF, BG, and MLP classifiers, exceeding all others including MFCC. It can also be observed that both FESF and WHFESF recognition results are overall better than other acoustic features commonly used in the current study, regardless of whether it is TORGO or UA Speech database. This indicates that the feature enhancement method significantly improves the classification accuracy, has the robustness to correctly recognize speech with different degrees of dysarthria, and can effectively improve the performance of VPD systems.

**Table 4.** Comparisons of statistical features presented by FEMD/WHFEMD with common acoustic features (UA Speech database).

| Type | Acc (%) | Type | Acc (%) | Type | Acc (%) |
|---|---|---|---|---|---|
| PSD.RF | 49.59 | PSD.BG | 50.49 | PSD.MLP | 51.15 |
| PITCH.RF | 50.74 | PITCH.BG | 54.03 | PITCH.MLP | 53.82 |
| LPC.RF | 55.92 | LPC.BG | 55.27 | LPC.MLP | 55.27 |
| LSP.RF | 52.80 | LSP.BG | 54.01 | LSP.MLP | 54.01 |
| FESF.RF | 58.37 | FESF.BG | 56.53 | FESF.MLP | 56.53 |
| WHFESF.RF | 63.99 | WHFESF.BG | 62.65 | WHFESF.MLP | 62.87 |

**Table 5.** Comparisons of statistical features presented by FEMD/WHFEMD with common acoustic features (TORGO database).

| Type | Acc (%) | Type | Acc (%) | Type | Acc (%) |
|---|---|---|---|---|---|
| PSD.RF | 77.26 | PSD.BG | 77.74 | PSD.MLP | 78.29 |
| PITCH.RF | 78.36 | PITCH.BG | 79.93 | PITCH.MLP | 78.77 |
| LPC.RF | 79.04 | LPC.BG | 77.96 | LPC.MLP | 77.74 |
| LSP.RF | 78.29 | LSP.BG | 78.15 | LSP.MLP | 78.15 |
| MFCC.RF | 80.75 | MFCC.RF | 79.32 | MFCC.MLP | 78.29 |
| FESF.RF | 79.45 | FESF.BG | 79.38 | FESF.MLP | 78.28 |
| WHFESF.RF | 81.10 | WHFESF.BG | 80.00 | WHFESF.MLP | 78.84 |

**Table 6.** Recognition accuracy of FEMD and WHFEMD applied to PSD/GFCC (UA Speech database).

| Type | Acc (%) | Type | Acc (%) | Type | Acc (%) |
|---|---|---|---|---|---|
| PSD.RF | 49.59 | PSD.BG | 50.49 | PSD.MLP | 51.15 |
| MFCC.RF | 67.10 | MFCC.BG | 65.10 | MFCC.MLP | 65.10 |
| FEPSD.RF | 60.45 | FEPSD.BG | 59.33 | FEPSD.MLP | 58.84 |
| WH2FEPSD.RF | 67.73 | WH2FEPSD.BG | 66.65 | WH2FEPSD.MLP | 66.65 |
| GFCC.RF | 76.54 | GFCC.BG | 73.25 | GFCC.MLP | 80.02 |
| WHGFCC.RF | 80.08 | WHGFCC.BG | 76.18 | WHGFCC.MLP | 82.50 |

In Table 6, the application of FEMD to extract PSD features (FEPSD) on the UA Speech database improved RF classification accuracy from 49.59% to 60.45%, BG from 50.49% to 59.33%, and MLP from 51.15% to 58.84%. After further WHFEMD enhancement (WH2FEPSD), the RF, BG, and MLP classifiers improved on PSD by 18.14%, 16.16%, and 15.5%, respectively. These improvements show the importance and feasibility of WHFEMD and FEMD methods on this feature.

It can also be obtained from the above gains when the WHFESF statistical features are used in combination with GFCC features (WHGFCC). The performance of GFCC will be improved from 76.65% to 80.08% on the RF classifier, from 73.25% to 76.18% using the BG classifier, and from 80.02% to 82.50% by MLP, which likewise suggests the adaptability of the WHFEMD method. Notably, all of them surpassed established acoustic features including MFCC on this database, outperforming existing methods [33].

**Table 7.** Recognition accuracy of FEMD and WHFEMD applied to PSD/GFCC (TORGO database).

| Type | Acc (%) | Type | Acc (%) | Type | Acc (%) |
| --- | --- | --- | --- | --- | --- |
| PSD.RF | 77.26 | PSD.BG | 77.74 | PSD.MLP | 78.29 |
| MFCC.RF | 80.75 | MFCC.BG | 79.32 | MFCC.MLP | 78.28 |
| FEPSD.RF | 79.52 | FEPSD.BG | 79.38 | FEPSD.MLP | 79.38 |
| WH2FEPSD.RF | 81.64 | WH2FEPSD.BG | 80.89 | WH2FEPSD.MLP | 80.89 |
| GFCC.RF | 85.14 | GFCC.BG | 83.63 | GFCC.MLP | 86.78 |
| WHGFCC.RF | 86.71 | WHGFCC.BG | 86.3 | WHGFCC.MLP | 90.58 |

Applying FEMD and WHFEMD to enhance PSD and GFCC on TORGO (as shown in Table 7) similarly improves their performance, with PSD improving by 4.38%, 3.15%, and 2.6%, respectively, and GFCC also showing relative increases of 1.57%, 2.67%, and 3.8%. The method proposed also gives more accurate results than the previous work on classification and recognition tasks using the same TORGO corpus [24]. In conclusion, these findings demonstrate the feature enhancement approach exhibits good feasibility and effectiveness in recognizing various dysarthria speech severity levels from normal speech, outperforming traditional prominent features. So, augmenting core acoustic representations via the non-linear signal decomposition techniques realizes discernible benefits.

In the following, the performance of GFCC and PSD on two databases will be further analyzed from the perspective of confusion matrix. And RF and BG are here taken as examples to compare the performance of WH2FEPSD and WHGFCC obtained by using the feature enhancement algorithms proposed applied on PSD and GFCC. For instance, "576" in Figure 4(a) indicates that the WHGFCC feature set extracted via WHFEMD correctly predicted 576 "severe" samples using Bagging on the UA Speech database.

As can be seen in Figure 4, the proposed approach outperformed all other methods for classification using the UA Speech database, which dominated in all categories in the experiments. Testing on the TORGO database produced one exception (Figure 5). For severe dysarthria, the Bagging classifier applying the augmented GFCC features reduced the number of correctly categorized data by 3 compared to GFCC alone. Nevertheless, the enhanced feature representation maintained robustness across differing severity levels of dysarthria (severe, moderate, mild) relative to the original features. It was also successful in decreasing classification confusion overall. These points suggest the method can comprehensively capture pathologic speech traits regardless of severity degree. Moreover, the single minor deviation may represent more of an isolated anomaly than true refutation.

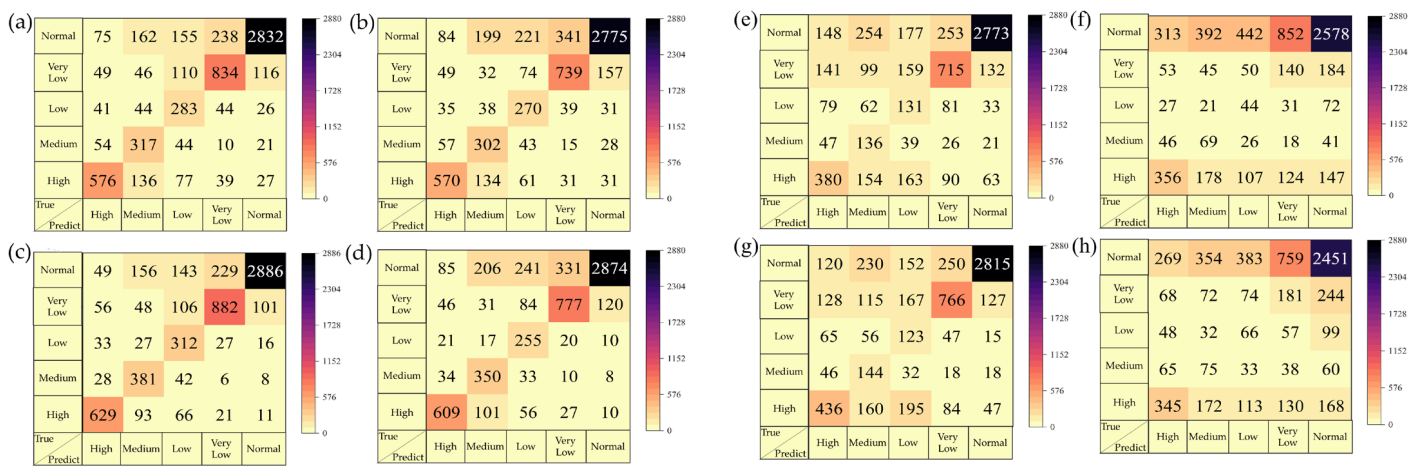

**Figure 4** Confusion matrix of WHGFCC and GFCC/WH2FEPSD and PSD features on UA Speech database: **(a)** WHGFCC+ Bagging; **(b)** GFCC+ Bagging; **(c)** WHGFCC+ RF; **(d)** GFCC+ RF ; **(e)** WH2FEPSD+ Bagging; **(f)** PSD+ Bagging; **(g)** WH2FEPSD+ RF; **(h)** PSD+ RF. (a-d) is the WHGFCC and GFCC features; (e-f) is WH2FEPSD and PSD features.

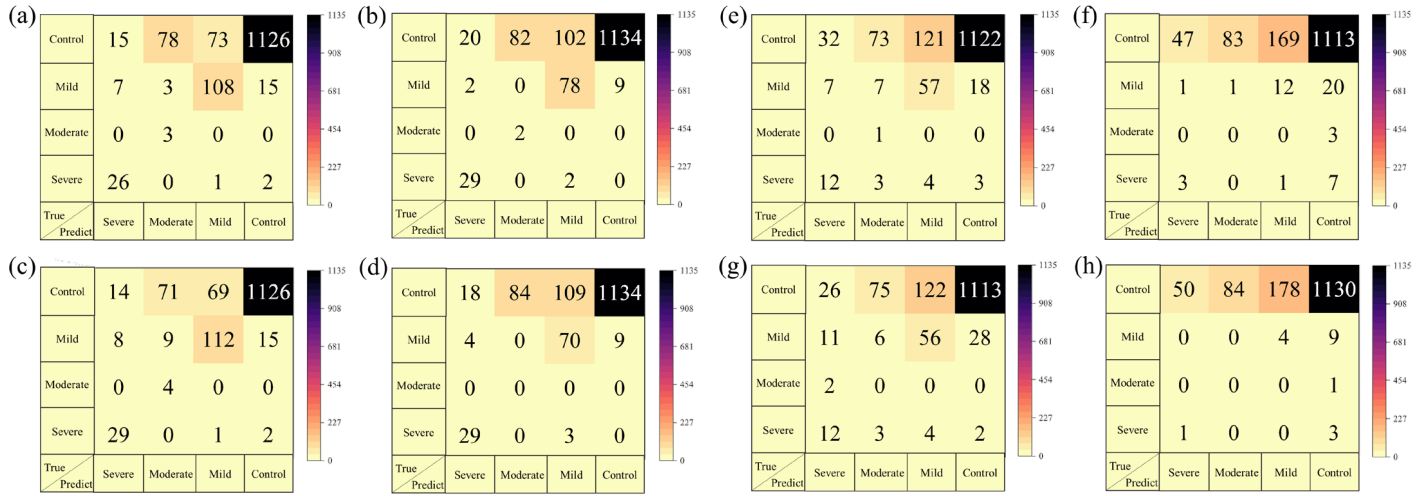

**Figure 5.** Confusion matrix of WHGFCC and GFCC/ WH2FEPSD and PSD features on TORGO database: **(a)** WHGFCC+ Bagging; **(b)** GFCC+ Bagging; **(c)** WHGFCC+ RF; **(d)** GFCC+ RF; **(e)** WH2FEPSD+ Bagging; **(f)** PSD+ Bagging; **(g)** WH2FEPSD+ RF; **(h)** PSD+ RF. (a-d) is the WHGFCC and GFCC features; (e-f) is WH2FEPSD and PSD features.

3.3.2. Comparison of complexity between WHFEMD and existing algorithms

**Table 8.** Comparison of complexity and runtime of WHFEMD and traditional algorithms.

| Algorithm | Parameters ($10^5$) | Model size (MB) | Runtime(s) | Time Complexity |
| --- | --- | --- | --- | --- |
| Adaptive filtering method | 1.02 | 4.38 | 482.56 | $O(N)$ |
| Spectral subtraction | 1.25 | 4.58 | 332.72 | $O(N+2N\log N)$ |
| Wavelet-based algorithm | 0.24 | 9.38 | 956.89 | $O(N+N\log N)$ |
| WHFEMD (ours) | 1.25 | 5.58 | 1048.49 | $O(N^2+N\log N)$ |

Traditional speech enhancement algorithms generally rely on adaptive filtering, spectral subtraction or wavelet transform-based methods. The basic adaptive filter method [34] realizes noise suppression by estimating statistical properties of the noise and the speech signal correlation to adjust the filter parameters. The basic spectral subtraction method [35] instead achieves this by calculating the speech spectrum and subtracting its noise. Wavelet packet-based speech enhancement approaches [36] utilize decomposition across wavelet packet coefficients at varied scales/frequency bands, attenuating noise components. As shown in Table 8, the proposed WHFEMD algorithm exhibits comparable parameter and memory sizes to these traditional methods, belonging to the same order of magnitude. However, increased runtime mainly stems from the iterative decomposition process of EMD. Crucially though, as per recognition accuracy results in experimental sections, WHFEMD augments discriminative ability for pathological speech by leveraging signal decomposition. This thereby improves the recognition ability. Therefore, WHFEMD is desirable for the feature representation of pathological speech features such as dysarthria. Its balanced optimization of efficiency and recognition fidelity merits.

Meanwhile, computational complexity analysis of different experiments provides useful context for comprehensively evaluating and comparing algorithm performance. Given this study focuses on proposing feature enhancement methods for dysarthric pathological speech, backend classifiers remain consistent across experiments. Therefore, comparative analyses principally center on complexity metrics relating to feature extraction algorithms. Parameter count offers insights into relative model scale. Runtime encompasses feature extraction and storage time for the entire speech corpora in both datasets.

As shown in Table 9, the statistical features of the algorithm proposed in this paper, FESF and WHFESF, are both parametric quantities of orders of magnitude in $10^5$, and the running times are 646s and 951s respectively. Combined with the results of the previous Table 3-Table 4 recognition accuracy results, the accuracy has improved, indicating that the algorithm is efficient and accurate for pathology speech recognition. The PSD feature enhancement (WH2FEPSD) of this paper, as a fusion feature of WHFESF and FEPSD, reduced total runtime versus the individual features by 560s while maintaining comparable dimensionality to FEPSD. Concurrently on TORGO and UA Speech databases, WH2FEPSD classification accuracy improved at minimum of 2.6% and 7.69% respectively (Table 6-Table 7). Similarly, WHGFCC - combining GFCC and WHFESF - decreased runtime 502s versus individual features, and raised recognition over 1.57%. Overall, though feature extraction parameters increased, the proposed algorithm markedly enhanced both runtime and recognition accuracy. This confirms salient representation learning while controlling computational overhead.

**Table 9.** Comparisons of complexity and runtime of different features.

| Features | Parameters ($10^5$) | Runtime(s) |
|---|---|---|
| FESF | 1.02 | 646 |
| WHFESF | 1.25 | 951 |
| PSD | 0.24 | 581 |
| FEPSD | 1.25 | 878 |
| WH2FEPSD | 1.26 | 1269 |
| GFCC | 0.25 | 620 |
| WHGFCC | 1.25 | 1068 |

3.3.3. Application of WHFEMD to unbalanced classification algorithms

As presented in Tables Table 4-Table 7, applying the proposed WHFEMD algorithm to GFCC features yielded optimal performance. So in this section, we will focus on applying the previously introduced unbalanced classification method [31] for categorizing WHGFCC feature inputs, to further evaluate both the proposed feature enhancement technique as well as the prior unbalanced classifiers.

The results in Table 10 demonstrate the proposed unbalanced classification approach based on Support Vector Machines and Naive Bayes outperforms the original SVM and NB models. Specifically, improvements of 12.18% and 45.15% were observed on the TORGO database, and 34.55% and 15.56% on the UA Speech database respectively. This confirms the unbalanced method notably surpasses utilizing either SVM or NB in isolation. In general, substantial gains were achieved compared to single classifier performance, as validated on both datasets. Additionally, when the proposed feature enhancement algorithm WHFEMD was applied to GFCC features, even better outcomes prevailed against Random Forest, Bayesian Network and Multi-Layer Perceptron benchmarks featured in Table 10. Taken together, these findings lend strong support to the promising potential of the method introduced in this study for classifying dysarthria impairments in speech. Overall, the framework shows early promise as an objective evaluation tool with importance for VPD systems.

**Table 10.** Comparison of identification accuracy rates on the TORGO and UA Speech databases.

| Algorithms | TORGO (%) | UA Speech (%) |
|---|---|---|
| SVM | 77.84 | 51.66 |
| Resample.SVM | 57.81 | 38.29 |
| Spreadsubsample.SVM | 60.22 | 34.60 |
| Smote.SVM | 56.02 | 43.30 |
| PCA.Smote.SVM | 89.13 | 86.24 |
| ProposedMethod.SVM | 90.02 | 86.21 |
| NB | 40.12 | 60.30 |
| Resample.NB | 39.57 | 58.90 |
| Spreadsubsample.NB | 38.69 | 58.29 |
| Smote.NB | 42.40 | 59.23 |
| PCA.Smote.NB | 57.47 | 59.36 |
| ProposedMethod.NB | 85.27 | 75.86 |

## 4. Conclusions

As a kind of speech disorder, patients will have slow and slurred pronunciation. The standard speech recognition system cannot meet the demand of VPD systems. There are two problems: the inherent nonlinear and non-stationary attributes of pathological speech and the necessity to capture feature parameters that encapsulate the pathological information of both the vocal tract and vocal folds. As a result, this study proposes a sophisticated feature representation algorithm designed to optimally represent the pathological speech properties and, in turn, augment model performance. This method, denoted as WHFEMD, unfolds in two principal phases. Initially, the FEMD approach commences with FFT processing of the raw signal, followed by the introduction of the EMD algorithm, which adaptively decomposes the signal into its IMFs. Subsequently, the FWHT is employed to refine the construction of these functions, yielding the final coefficients. The ensuing experiments involve the effectiveness evaluation of statistical features and the performance of applications on PSD and GFCC features. In this paper, RF, Bagging and MLP are used as classifiers for evaluation on TORGO and UA Speech databases. The results demonstrate that the statistical features extracted by the WHFEMD and FEMD algorithms facilitate elevated recognition accuracy for dysarthric speech and significantly bolster the performance of PSD and GFCC analyses. The granular feature information pertaining to the local characteristics of pathological speech can be extracted via EMD decomposition. This means that our method permits a more profound investigation into the vocal mechanisms underlying pathological speech and promises to yield auxiliary insights for clinical diagnostics and therapeutic strategies. The integration of FWHT is instrumental in extracting features related to the pathological conditions of the vocal tract and vocal folds, thereby enhancing efficiency—a critical factor for handling the voluminous data inherent in clinical practice.

In the future, the feasibility of the dysarthria severity assessment model will be investigated by combining multi-modal data such as acoustic, vocal folds, and facial micro expressions, to further explore the etiology of dysarthria in terms of human articulatory mechanisms.

**Author Contributions:** Conceptualization, T.Z. and C.D.; methodology, T.Z.; software, T.Z.; validation, C.D., S.D., and H.L; formal analysis, W.Z., H.L., H. J. and S.D.; data curation, S.D.; writing—original draft preparation, T.Z.; writing—review and editing, T.Z., H. J. and S.D.; visualization, T.Z.; supervision, S.D and H.L.; project administration, S.D.; funding acquisition, S.D.. All authors have read and agreed to the published version of the manuscript.

**Funding:** This research was funded by National Natural Science Foundation of China Youth Science Foundation, under Grant No. 12004275, Applied Basic Research Program of Shanxi Province on the surface of natural funds, grant number 20210302123186.

**Institutional Review Board Statement:** Not applicable.

**Informed Consent Statement:** Not applicable.

**Data Availability Statement:** The TOGRO database is available online: https://www.cs.toronto.edu/~complingweb/data/TORGO/torgo.html. And the UA Speech database is also public, you can find it at http://www.isle.illinois.edu/sst/data/UASpeech/.

**Conflicts of Interest:** The authors declare no conflicts of interest.